# Uma estratégia para recomendação de especialistas a partir de dados abertos disponíveis na Plataforma Lattes

Sérgio José de Sousa (Centro Federal de Educação Tecnológica de Minas Gerais – CEFET-MG, MG, Brasil) – sergio7sjs@gmail.com

Thiago Magela Rodrigues Dias (Centro Federal de Educação Tecnológica de Minas Gerais – CEFET-MG, MG, Brasil) – thiago@div.cefetmg.br

Adilson Luiz Pinto (Universidade Federal de Santa Catarina – UFSC, SC, Brasil) – adilson@cin.ufsc.br

**Resumo.** Com o crescente volume de dados e usuários de sistemas de currículos, a dificuldade de encontrar especialistas é cada vez maior. Neste trabalho é proposto metodologia de extração de dados abertos dos currículos da Plataforma Lattes, um tratamento para esses dados e investiga uma abordagem de Agente de Recomendação baseado em redes neurais profundas com autoencoder.

**Palavras-chave**: Recomendação de Especialista; Plataforma Lattes; Recuperação de Informação; Redes Neurais; Deep Learning.

## 1. INTRODUÇÃO

Com o crescente volume de dados é um desafio cada vez maior encontrar informações que se desejam. Assim como ocorre com diversas áreas, a busca por especialistas possui pouco suporte e poucos recursos de busca, servido apenas como auxilio não quantitativos.

Assim como o aumento no número de comunidades, o aumento no número de pesquisadores disponíveis também é visível, sendo na Plataforma Lattes mais de 6 milhões de indivíduos registrados. Logo, realizar buscar não automatizadas pode ser extremamente custoso.

Em Dias e Moita (2015) os autores mostram que a Plataforma Lattes do Conselho Nacional de Desenvolvimento Científico e Tecnológico – CNPq, possui importantes e relevantes dados para a compreensão dos pesquisadores e da ciência brasileira, incluindo informações pessoais, profissionais e acadêmicas, com a produção científica e tecnológica. Portanto se torna uma importante fonte de informações desses indivíduos. Mesmo com tamanha relevância esse repositório não foi ainda amplamente explorado (FERRAZ, QUONIAM e MACCARI, 2014)

Neste contexto, agentes de recomendação e técnicas de ranqueamento são considerados boas soluções para o problema da sobrecarga de informação, tentando extrair através de tópicos de interesse possibilidades para aquele problema. A recomendação de especialistas e recuperação de especialistas são subcampos que visam recomendar aos usuários uma lista de especialistas que podem possuir o conhecimento específico procurado, facilitando colaborações e comunicação através dessa recomendação, sendo parte fundamental na gestão do conhecimento (SERDYUKOV et al, 2008).

Neste trabalho é proposto uma técnica para extração de radicais textuais que possam descrever as especialidades em cada currículo e um agente de recomendação baseado em aprendizado profundo, utilizando de redes neurais artificiais com várias camadas capaz de aprender vínculos contextuais e aprender relações semânticas do especialista com suas especialidades. Para no fim, proporcionar uma ferramenta capaz de formar novos grupos de pesquisas, recomendar avaliadores para periódicos ou pesquisadores no geral, tendo como base os currículos da Plataforma Lattes.

## 2. TRABALHOS RELACIONADOS

Em Cunha et al. (2013) foi analisada as redes semânticas dos termos extraídos dos artigos publicados na Nature de 1999 a 2008 com objetivo de verificar o relacionamento temporal em uma comunidade científica. Através de diversas métricas e indicadores bibliométricos foi verificado um fenômeno de mundo pequeno (WATTS e STROGATZ, 1998) levando a supor que se tem uma alta probabilidade de duas palavras ligadas a uma terceira serem ligadas entre si formando uma clique.

No trabalho realizado por Gomes, Dias e Moita (2018) os principais tópicos de pesquisa dos doutores brasileiros tendo como fonte de dados as palavras chaves cadastradas nos artigos de periódicos e de anais em congressos registrados na Plataforma Lattes são analisados. É evidenciado a importância das palavras chaves que são apropriadas para representar temas de pesquisas, já nos títulos nem sempre é possível expressar todo conteúdo do trabalho. Por fim, os resultados mostram uma visão geral sobre os principais tópicos de pesquisa ao longo de 55 anos, ajudando a entender sobre como a ciência brasileira vem se desenvolvendo.

Em Trucolo e Digiampietri (2014a) é feita uma análise de tendência da produção científica brasileira dos programas de pós-graduação na área de ciências da computação, neste trabalho foram analisados 57.501 títulos de publicações distribuídos entre 45 programas entre os anos de 1991 e 2011. Foi quantificado os termos dos títulos e utilizada a medida TF-IDF que calcula a importância de um termo de um documento em relação a um conjunto de documentos, e por fim calculando regressões em cima dos valores encontrados. Um trabalho similar, mas focado nas produções científicas de doutores da Ciência da Informação pode ser visto em Trucolo e Digiampietri (2014b) conseguindo identificar tendências de pesquisas também nessa área.

Os periódicos científicos são uma das principais fontes de conhecimento científico, Juliani e Martignago (2018) destacam que, apesar dos avanços tecnológicos o tempo gasto desde a submissão até a publicação de manuscritos continua alto. Em uma pesquisa em periódicos da Ciência da Informação publicados em 2015, 31% desses levam mais de um ano para se publicar um trabalho. Um dos motivos dessa demora está atrelado ao processo de avaliação, incluindo o tempo de se encontrar um avaliador especialista no assunto do artigo. Os autores então propuseram uma ferramenta que indexou 95 periódicos de Administração, 30 de Ciência da Informação e 94 da Educação, extraiu-se os resumos de mais de 95 mil artigos e com base em um índice de similaridade dos resumos são sugeridos possíveis avaliadores.

Mesmo com resultados significativos, um estudo abrangente com grandes repositórios de dados abertos, para a identificação de especialistas ainda não foi proposto, tendo em vista os desafios e limitações de processamento ou fonte de dados existentes. Logo, diferentemente dos trabalhos anteriores, este estudo visa analisar grandes repositórios de dados abertos, para, baseado em informações de sua produção científica, realizar a recomenda de especialista.

## 3. METODOLOGIA

As subseções (3.1) Obtenção dos dados, (3.2) Tratamentos dos dados e (3.3) Agente de recomendação, tratam sobre a estrutura completa que se propõe este trabalho que pode ser vista de maneira resumida na Figura 1.

### 3.1 Obtenção dos dados

O Currículo Lattes é o padrão nacional adotado pela maioria das instituições de fomentos e usada para registrar a vida acadêmica de estudantes e pesquisadores. Além de integrar todas as áreas de conhecimento e incluir em seus registros artigos publicados em periódicos e congressos nacionais que muitas vezes não são indexados. Por essas qualidades e por ainda não ter sido inteiramente explorado a Plataforma Lattes foi escolhida como fonte de dados.

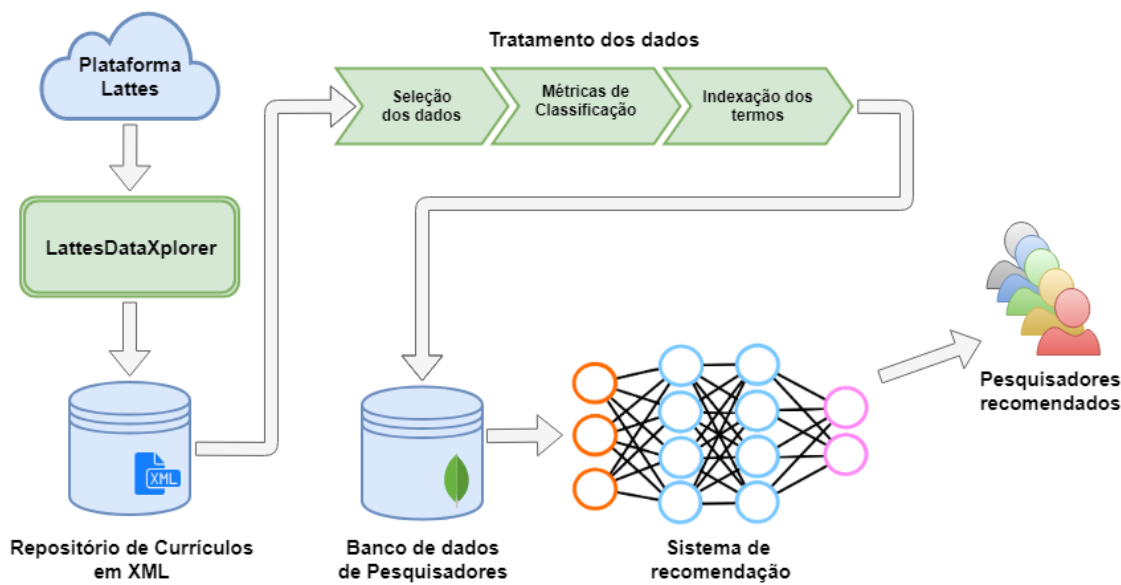

**Figura 1 – Arquitetura da Plataforma de Extração e Integração**
**Fonte: Próprio autor**

Esses dados estão disponíveis livremente em uma interface de visualização que não exibe todas as informações cadastradas nos currículos. As consultas são realizadas individualmente o que é muito custoso de se realizar manualmente, sendo então necessário o uso técnicas e ferramentas para essa extração. Para tanto, foi utilizado o LattesDataXplorer, um arcabouço desenvolvido para extração, tratamento e análise de dados da Plataforma Lattes (DIAS, 2016).

### 3.2 Tratamento dos dados

Com currículos armazenados localmente inicia-se então a etapa de tratamento dos dados, extraindo títulos juntamente com suas palavras chaves de cada trabalho publicados em anais de congressos ou em periódicos, e ainda, títulos de projetos. Após, algumas etapas de transformação são aplicadas: (1) identificar o idioma dos títulos e palavras chaves; (2) transformar todos os caracteres em minúsculo; (3) aplicar remoção de palavras vazias (stopwords), termos que agregam pouco valor semântico por serem muito comuns ou serem palavras funcionais (como, o, a, em, no).

Após a separação dos dados inicia-se as análises com algumas métricas de quantificação e classificação dos termos, para isso utilizou-se técnicas de radicalização (stemming) nas palavras, o que consiste em reduzir as palavras em seus radicais, permitindo agrupar palavras que antes poderiam ser classificada como distintas como por exemplo, os termos "bibliometria" e "bibliométrico" tratam de um mesmo assunto, permitindo ao aplicar o filtro de stopwords e de radicalização obter o radical "bibliometr" para ambos os casos.

Por fim é possível criar um dicionário de radicais contendo a frequência de cada palavra juntamente com a medida estatística TF-IDF para que ambos sejam utilizados nos agentes de recomendação.

### 3.3 Tratamento dos dados

Com auxílio de arquiteturas de redes neurais profundas o agente de recomendação é capaz de dado um conjunto de palavras dos quais se espera que o especialista tenha conhecimento ele retorne possibilidades de indivíduos juntamente com sua classificação. Para isso é necessário encontrar uma arquitetura ideal para o problema proposto, quantidade de camadas e neurônios em cada uma, bem como os principais hiperparâmetros (HE et al 2017; WEI et al., 2017).

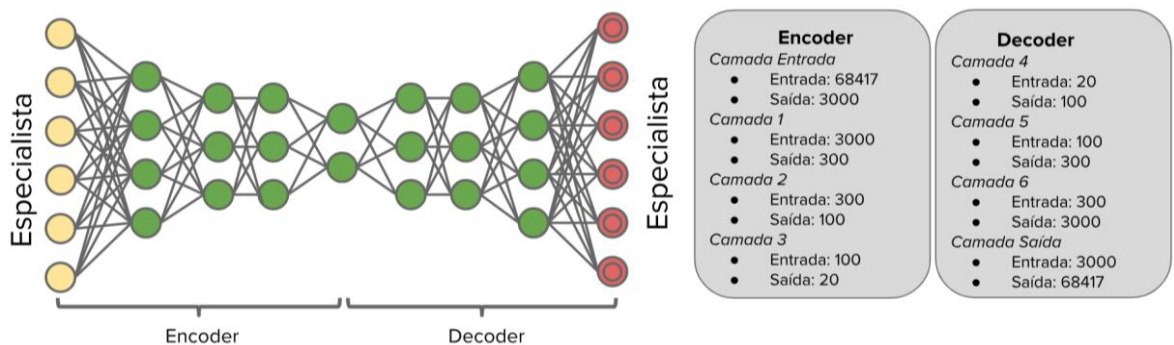

**Figura 2 – Modelo autoencoder proposto**
Fonte: Próprio autor

A Figura 2 exibe o modelo utilizado inspirado no tralho de Huang et al. (2013). Onde temos um autoencoder (HINTON e SALAKHUTDINOV, 2006) com o propósito de reduzir dimensionalmente os dados, conseguindo uma representação mais compacta e com mais valor semântico. O modelo possui 4 camadas na etapa de Encoder e 4 na etapa de Decoder, com as quantidades de entrada e saída descritas na figura. Foi utilizada a função de ativação LeakyReLU (XU, et al. 2015), uma extensão da função ReLU que obteve melhores resultados. Como função de perda foi utilizada a Similaridade de cosseno que verifica o quão similar o tensor de entrada é o tensor de saída. Por último a função de otimização Adam (KINGMA, 2014) com taxa de aprendizagem 0,00001.

Após treinado, aproveita-se a etapa de *Encoder* para redução de dimensão, saindo de 68.417 para um espaço de 20 dimensões, com isso, aplica-se em todos os especialistas gerando um novo vetor de características para cada. Então, ao realizar uma busca basta indexa-las com os métodos descritos anteriormente, transforma-la utilizando o *Encoder* e por fim calcular a similaridade do cosseno entre a busca e os especialistas, ordenando pela maior similaridade.

## 4. RESULTADOS

O quantitativo dos dados a serem analisados são apresentados na Tabela 1. Estes dados extraídos dos currículos cadastrados na Plataforma Lattes de indivíduos que possuem doutorado concluído como maior titulação acadêmica, representam uma parcela significativa do total de registro sobre produção de artigos em periódicos (75%) e de artigos em anais de congressos (63%).

**Tabela 1**: Quantitativo dos dados analisados

| Tipo de Dado | Quantidade |
|---|---|
| Artigos em Anais de Congressos | 12.456.432 |
| Artigos em Periódicos | 6.897.234 |
| Projetos de Pesquisa | 234.897 |
| Projetos de Extensão | 198.345 |
| Termos Extraídos e Radicalizados | 68.417 |
| Utilizações dos termos | 193.440.564 |

Os dados foram coletados em janeiro de 2018, totalizando 308.256 currículos de indivíduos com doutorado concluído. Para as análises, foram considerados os artigos publicados em anais de congressos e em periódicos referentes ao período de 1962 até 2018.

De posse destes dados, todos os títulos passaram pelo processo de transformação, conforme descrito na seção 3.2 para que o agente de recomendação possa ser aplicado aos termos identificados.

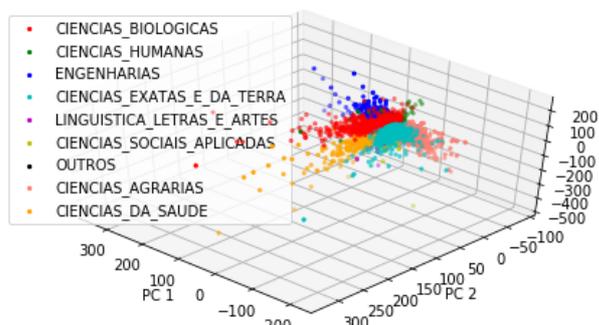

**Figura 3 – Visualização 3D do novo espaço gerado**
**Fonte: Próprio autor**

O modelo foi treinado por 18 épocas, levando cerca de 200 minutos em cada, atingindo o erro mínimo de reconstrução de 0,33. Ao realizar todo o procedimento descrito no item 3.3 podemos gerar um gráfico dos resultados, aplicando PCA para transformar o espaço de dimensão 20 para 3 e assim podermos visualizar o resultado. Na Figura 3 podemos ver o início de formações de grupos onde cada ponto representa um especialista e a cor a sua grande área, indicando que o modelo conseguiu aprender através das palavras contidas nos títulos uma nova representação para os dados.

## 5. CONSIDERAÇÕES FINAIS

Para os testes que estão sendo realizados são considerados os currículos dos doutores cadastrados na Plataforma Lattes até janeiro de 2019 totalizando 308.256 indivíduos de diversas áreas do conhecimento. Essa parcela dos dados é onde estão concentradas o maior volume de informações pois cerca de 70% do total de artigos registrados na plataforma são de autoria de doutores (DIAS, 2016).

Diante disso, a pesar de ser um conjunto pequeno de indivíduos, representam uma significativa parcela da produção científica brasileira disponibilizada em acesso aberto. Os resultados iniciais realizados com os agentes implementados possuem uma boa performance computacional, sendo processado um grande volume de dados em tempo hábil.

No momento atual da pesquisa estão sendo avaliados os resultados obtidos já implementados para identificar a eficácia dos resultados com o intuito de propor melhorias nas recomendações obtidas.

**REFERÊNCIAS**